\documentclass[a4paper,12pt]{article}
\usepackage{amsfonts,amsthm,amsmath,amssymb,graphicx,hyperref,color,youngtab}

\newcommand{\be}{\begin{equation}}
\newcommand{\bea}{\begin{eqnarray}}

\newcommand{\ee}{\end{equation}}
\newcommand{\eea}{\end{eqnarray}}

\begin{document}

\makeatletter
\@addtoreset{equation}{section}
\makeatother
\renewcommand{\theequation}{\thesection.\arabic{equation}}

\rightline{}
\vspace{1.8truecm}

\vspace{15pt}


{\LARGE{  
\centerline{\bf From Gauss Graphs to Giants} 
}}  

\vskip.5cm 

\thispagestyle{empty}
    {\large \bf 
\centerline{Robert de Mello Koch$^{a,b,}$\footnote{ {\tt robert@neo.phys.wits.ac.za}} and
Lwazi Nkumane$^{b,}$\footnote{ {\tt lwazi.nkumane@gmail.com}}}}

\vspace{.4cm}
\centerline{{\it ${}^a$ School of Physics and Telecommunication Engineering},}
\centerline{{ \it South China Normal University, Guangzhou 510006, China}}

\vspace{.4cm}
\centerline{{\it ${}^b$ National Institute for Theoretical Physics,}}
\centerline{{\it School of Physics and Mandelstam Institute for Theoretical Physics,}}
\centerline{{\it University of the Witwatersrand, Wits, 2050, } }
\centerline{{\it South Africa } }

\vspace{1.4truecm}

\thispagestyle{empty}

\centerline{\bf ABSTRACT}

\vskip.4cm 
We identify the operators in ${\cal N}=4$ super Yang-Mills theory that correspond to ${1\over 8}$-BPS giant gravitons
in AdS$_5\times$S$^5$.
Our evidence for the identification comes from (1) counting these operators and showing agreement with 
independent counts of the number of giant graviton states, and (2) by demonstrating a correspondence between 
correlation functions of the super Yang-Mills operators and overlaps of the giant graviton wave functions. 

\setcounter{page}{0}
\setcounter{tocdepth}{2}

\newpage

\tableofcontents

\setcounter{footnote}{0}

\linespread{1.1}
\parskip 4pt

{}~
{}~

\section{Introduction}\label{intro}

The AdS/CFT correspondence\cite{Maldacena:1997re} provides a beautiful realization of 't Hooft's proposal that the large 
$N$ limit of Yang-Mills theories are equivalent to string theory\cite{'tHooft:1973jz}. 
Most studies of the correspondence have focused on the planar limit, which holds classical operator dimensions fixed as 
we take $N\to\infty$.
There are non-planar large $N$ limits of the theory \cite{Balasubramanian:2001nh}, which are defined by considering
operators with a bare dimension that is allowed to scale with $N$ as we take $N\to\infty$.
These limits are relevant for the AdS/CFT correspondence.
The limit on which we will focus in this study considers operators with a dimension that scales as $N$.
Our focus is on operators relevant for the description of giant graviton branes\cite{McGreevy:2000cw,Hashimoto:2000zp,Grisaru:2000zn}. 

The worldvolume of the most general ${1\over 8}$-BPS giant graviton can be described as the intersection
of a holomorphic complex surface in $\mathbb{C}^3$ with the five sphere $S^5$ of the AdS$_5\times$S$^5$
spacetime\cite{Mikhailov:2000ya}.
It is possible to quantize these giant graviton configurations and then to count them\cite{Biswas:2006tj}. 
Remarkably, this quantization leads to the Hilbert Space of $N$ noninteracting Bose particles in a 3d harmonic 
oscillator potential, a result conjectured in \cite{Beasley:2002xv}.
In \cite{Mandal:2006tk} ${1\over 8}$-BPS states which carry three independent angular momenta on $S^5$
were counted. 
This counting problem can again be mapped to counting energy eigenstates of a system of $N$ bosons 
in a 3-dimensional harmonic oscillator.
Both of these analysis \cite{Biswas:2006tj,Mandal:2006tk} make use of a world volume description of the branes.
Finally, an index to count single trace BPS operators operators has been 
constructed \cite{Kinney:2005ej,Romelsberger:2005eg}.
The index has been computed both at weak coupling (using the gauge theory) and at strong coupling (as a sum over 
the spectrum of free massless particles in AdS$_5\times$S$^5$) and the results again agree with \cite{Biswas:2006tj,Mandal:2006tk}.

Given the AdS/CFT correspondence, this counting should also arise in the dual ${\cal N}=4$ super Yang-Mills theory,
when the operators of a bare dimension of order $N$ and vanishing anomalous dimension are considered.
One of our goals in this study is to demonstrate this.

A crucial ingredient in the study of operators with a bare dimension of order $N$, has been the construction of bases of
operators developed in \cite{Corley:2001zk,Balasubramanian:2004nb,Brown:2007xh,Kimura:2007wy,Kimura:2008ac,Bhattacharyya:2008rb,Bhattacharyya:2008xy,Brown:2008ij}.
These bases diagonalize the free field theory two point function to all order in $1/N$ and mix weakly when the Yang-Mills
coupling is switched on.
Using these bases as a starting point, the spectrum of anomalous dimensions for a class operators of bare dimension of 
order $N$ has been computed in \cite{Carlson:2011hy,Koch:2011hb,deMelloKoch:2012ck}.
The operators are constructed using the three complex adjoint scalars $Z$, $Y$ and $X$.
We use $n$ $Z$s, $m$ $Y$s and $p$ $X$s, fixing $n\sim N$ and $m,p\ll n$.
This implies that we are focusing on small deformations of ${1\over 2}$-BPS giant gravitons.
The operators of a definite scaling dimension are labeled by a permutation $\sigma\in S_m\times S_p$ and a pair of
Young diagrams $R\vdash n+m+p$ and $r\vdash n$.
The explicit form of these operators is
\bea
  O_{R,r}^{\vec m,\vec p}(\sigma)
  &=&{|H_X\times H_Y|\over \sqrt{p! m!}}\sum_{j,k}\sum_{s\vdash m}\sum_{t\vdash p}
\sum_{\vec\mu_1,\vec\mu_2}\sqrt{d_s d_t}
  \Gamma^{(s,t)}_{jk}(\sigma )\cr
&&\qquad\times B^{(s,t)\to 1_{H_X\times H_Y}}_{j \vec\mu_1}
B^{(s,t)\to 1_{H_X\times H_Y}}_{k \vec\mu_2} O_{R,(r,s,t)\vec\mu_1\vec\mu_2}
  \label{ggo}
\eea
The Young diagrams $R$ and $r$ both have $q$ rows for operators dual to a state of $q$ giant gravitons.
Each box in $R$ is associated with one of the complex fields, so that we can talk of a box as being a $Z$ box, a $Y$ box
or an $X$ box.
$r$ collects all of the $Z$ boxes.
The difference in the row length of the $q$th row in $R$ and $q$th row in $r$ is equal to the number of $X$s ($=p_q$)
and $Y$s ($=m_q$) in row $q$, so that $R_q-r_q=m_q+p_q$.
The right most boxes are $X$ boxes, the left most boxes $Z$ boxes and the $Y$ boxes are sandwiched in the middle.
The $q$ dimensional vector $\vec m$ collects the $m_i$, while $\vec p$ collects the $p_i$. 
The branching coefficients $B^{(s,t)\to 1_{H_Y\times H_X}}_{j\vec\mu}$ resolve the operator that 
projects from $(s,t)$, with $s\vdash m$, $t\vdash p$, an irreducible
representation  of $S_m\times S_p$, to the trivial (identity) representation of 
the product group $H_Y\times H_X$ with $H_Y=S_{m_1}\times S_{m_2}\times \cdots S_{m_q}$ and
$H_X=S_{p_1}\times S_{p_2}\times \cdots S_{p_q}$, i.e.
\bea
{1\over |H_X\times H_Y|}\sum_{ \gamma \in H_X\times H_Y } \Gamma^{(s,t)}_{ik} ( \gamma )
=\sum_{\vec\mu} 
B^{(s,t) \rightarrow 1_{H_X\times H_Y}}_{ i \vec\mu}  
B^{(s,t) \rightarrow 1_{H_X\times H_Y}}_{ k \vec\mu}
\eea
The operators $O_{R,(r,s,t)\vec\mu_1\vec\mu_2}$ are normalized versions of the restricted
Schur polynomials \cite{Bhattacharyya:2008rb}
\bea
\chi_{R,(r,s,t)\vec\mu_1\vec\mu_2}(Z,X,Y)=
{1\over n!m!p!}\sum_{\sigma\in S_{n+m+p}}\chi_{R,(r,s,t)\vec\mu_1\vec\mu_2}(\sigma)
{\rm Tr}(\sigma Z^{\otimes n}Y^{\otimes m}X^{\otimes p}),
\eea
which themselves provide a basis for the gauge invariant operators of the theory.
The restricted characters $\chi_{R,(r,s,t)\vec\mu_1\vec\mu_2}(\sigma)$ are defined by tracing the matrix representing
group element $\sigma$ in representation $R$ over the subspace giving an irreducible representation $(r,s,t)$ of the subgroup
$S_n\times S_m\times S_p$.
There is more than one choice for this subspace and the multiplicity labels $\vec\mu_1\vec\mu_2$ resolve
this ambiguity.
The operators $O_{R,(r,s,t)\vec\mu_1\vec\mu_2}$ given by
\bea
   O_{R,(r,s,t)\vec\mu_1\vec\mu_2} = \sqrt{{\rm hooks}_r{\rm hooks}_s{\rm hooks}_t\over {\rm hooks}_R f_R}
\, \chi_{R,(r,s,t)\vec\mu_1\vec\mu_2}
\eea
have unit two point function.
Although the definition of the Gauss graph operators $O_{R,r}(\sigma)$ is technically rather involved, they have
a very natural and simple interpretation in terms of the dual giant graviton branes plus open string excitations.
A Gauss graph operator that is labeled by a Young diagram $R$ that has $q$ rows corresponds to a system
of $q$ giant gravitons. 
The $Y$ and $X$ fields describe the open string excitations of the giants.
Each such field corresponds to a directed edge, an open string, which can end on any two (not necessarily distinct) of 
the $q$ branes.
The permutation $\sigma\in S_m\times S_p$ is a label which tells us precisely how the $m$ $Y$'s and the $p$ $X$'s
are draped between the $q$ giant gravitons.
The picture of directed edges stretched between $q$ dots is highly suggestive of a brane plus open string system, as
reflected in our language.
This interpretation is further supported by the fact that the only configurations that appear have the same number of 
strings starting or terminating on any given giant.
This nicely implements the Gauss Law of the brane world volume theory implied by the fact that the giant 
graviton has a compact world volume.
The Gauss graph operators which correspond to BPS states have all open strings described by loops that start at a
given giant and loop back to the same giant, i.e. no open strings stretch between giants.
In this case, we simply need to specify which brane the open string belongs to and this is most conveniently done
by partially labeling Young diagram $R$: in each box we place a $z$, an $x$ or a $y$.
Each row in the operator consists mainly of $Z$ fields, corresponding to the fact that the unexcited giant graviton
is dual to a half-BPS operator built only from $Z$s. 
The number of $x$ and $y$ boxes in a given row tell us how many $X$ and $Y$ strings attach to the corresponding
giant.

In the next section we will show the counting of these BPS states agrees with the counting 
of \cite{Biswas:2006tj,Mandal:2006tk}.
Motivated by this observation, we explore the link between the $N$ particle description employing the 3d harmonic oscillator
and the super Yang-Mills operators in section \ref{states}.
Our results shed light on the attractive possibility of an $N$ particle description of multi matrix models, suggesting that 
there maybe an extension of the famous free fermion/eigenvalue description of single matrix models \cite{Brezin:1977sv}.
Finally, we refer the reader to \cite{SanJurg} and \cite{Jeff} for further related background dealing with BPS giant gravitons
and to \cite{Bianchi:2003wx,Brown:2010pb,Pasukonis:2010rv} for further background relevant for the counting and
construction of ${1\over 4}$ and ${1\over 8}$-BPS operators for the regime where operator dimensions are less than $N$.

\section{Counting}\label{count}

As discussed in the introduction, our description of ${1\over 8}$-BPS operators is in terms of a Young diagram $R$
with partially labeled boxes.
When the boxes corresponding to $Y$ and $X$ fields are removed from the rows of $R$, we are left with the valid Young
diagram $r$.
An example of a valid ${1\over 8}$-BPS operator is
\bea
\young(zzzzzzzzyyx,zzzzzzzyx,zzzyy)\label{YDLABEL}
\eea
The boxes with label $z$ belong to the Young diagram $r$ and the boxes with label $y$ or $x$ are the ones that are 
removed from the Young diagram $R$ to obtain $r$.
The  operator labeled by the Young diagram shown in (\ref{YDLABEL}) corresponds to a system of 3 giant gravitons, 
with 2 $Y$ strings and an $X$ string attached to the first giant, a $Y$ and an $X$ string attached to the second giant 
and 2 $Y$ strings attached to the third giant.  
This description in terms of Gauss graph operators is valid in the case where $n$ the total number of boxes of the 
Young diagram $r$ and $m+p$ the total  number of the boxes that are added to the Young diagram $r$ to form $R$, 
are both large and of order $N\gg1$. 
In addition, $m+p\ll n$ and the number of rows of the Young diagram $R$ is of order $1=N^0$. 
Finally, the length of any row of $R$ is of order $N$, as is the difference between the length of any two
consecutive rows.

Let us first start by fixing our notation. 
We will denote by $R_i$ the number of boxes in the $i$th row of $R$, and we will denote by $m_i$ and $p_i$ the number 
of $Y$ and $X$ boxes to be removed from the $i$th row of $R$ to obtain $r$. 
Furthermore, $q$ will stand for the number of rows of $R$, $n$ will stand for the total number of boxes of $r$,
$m =\sum_{i=1}^q m_i$ and $p=\sum_{i=1}^q p_i$.
Hence, the total number of boxes of $R$ is then $n+m+p$. 
If we denote by $r_i$ the number of boxes in the $i$th row of $r$, then we have
$$
r_i = R_i - m_i - p_i
$$
In our conventions, we start the numbering of rows from top to bottom.
As already mentioned above, this description of ${1\over 8}$-BPS states is proved to work\cite{Koch:2011hb} in the cases that
\bea
R_i \sim N\qquad R_{i+1}- R_i \gg m+p \sim N \qquad q \sim N^0
\eea
We call this the displaced corners approximation because the neighboring corners of $R$ are separated by a huge 
number of columns. 
Outside this regime, things are more complicated and it is not even known if partially labeled Young diagrams 
can be used to describe these ${1\over 8}$-BPS states. 
The number of ${1\over 8}$-BPS operators is the same as the number of possible pairs $(R; r)$ counted with
multiplicity equal to the number of ways of assigning a valid vector $\vec{m}=(m_1,m_2,\dots,m_q)$.
Note that once the pair $(R; r)$ and the vector $\vec m$ are given, the vector $\vec p=(p_1,p_2,\dots,p_q)$ is
determined.
The first step towards counting the number of Gauss graph operators entails writing a generating function for the
number of pairs $(R;r)$. 
Our starting point is the observation that the Young diagrams are in one to one correspondence with partitions of integers. 
The generating function of the latter is given by
\bea
Z =\prod_{n=1}^\infty {1\over 1 - q^n}=\sum_{k=0}^\infty D_k q^k
\eea
where $D_k$ is the number of possible ways to partition an integer $k$.
This counting is too coarse for us to reach our goals: we need to track the number of parts in the partition which corresponds
to the number of rows in the Young diagram.
Indeed, we must encode the information about $q$ the number of rows of $R$, as well as the information about the 
different possible $m_i$'s and $p_i$'s in such a partition, to ensure that we are counting states in the regime in which
the Gauss graph operators provide a trustworthy description.
Both modifications are easy to take into account in our case of interest where $m_i+p_i+r_i\ll m_{i+1}+p_{i+1}+ r_{i+1}$
for all values of $i=1,2,\dots,q$.
The number of ways to partition an integer $k$ is given by the number of solutions to the equation
\bea
   k=\sum_i \chi_i n_i\qquad n_1\ge n_2\ge\cdots >0\qquad \chi_i\ge 0
\eea
Notice that the term $\chi_i n_i$ in the above equation is associated to the term  $(q^{n_i})^{\chi_i}$ in the 
expansion of $Z$. This term appears in the expansion for the term $(1 - q^{n_i})^{-1}$. 
Clearly then, to keep track of contributions from different rows $\chi_i$ we just need to multiply $q^n$ by an extra 
parameter $\chi$ and track the power of $\chi$.
So, we consider the following modification of the partition function $Z$
\bea
Z =\prod_{i=1}^\infty{1\over 1 -\chi q^n}=\sum_{k,d=0}^\infty
D_{k ; d} \chi^d q^k\label{pf}
\eea
where $D_{k ; d}$ counts the number of Young diagrams with $k$ boxes and $d$ rows. 
Next consider the information associated to the $m_i$'s and $p_i$'s. 
There is a potential complication because we want both $R$ and $r$ to be Young diagrams. 
However, in the displaced corners limit, we can ensure that this is not an issue.
Indeed, by taking $m,p\ll |r_{i+1}-r_i|$ for all $i$, we ensure that we can never pile enough $Y$ and $X$ boxes
onto a row to make it longer than the row above it.
Thus, we may treat the $m_i$'s  and $p_i$'s as independent, except for the requirement that $\sum_{i=1}^q m_i = m$ 
and $\sum_{i=1}^q p_i = p$. 
In terms of the partition function $Z$, this is equivalent to associating to each term $q^{\chi_i n_i}$,
a term $p^{b_i m_i}r^{c_i p_i}$ , where $b_i \ne \chi_i$ and $c_i\ne \chi_i$ in general. 
The latter condition is equivalent to associating the term $p^l r^m$, with $l,m= 0 , 1,\dots$ for each term $q^n$ in 
the product form of $Z$ in equation (\ref{pf}). 
Thus, we finally obtain the generating function
\bea
Z =\prod_{l=0}^\infty\prod_{m=0}^\infty\prod_{n=0}^\infty
{1\over 1 -\chi p^l r^m q^n}=
\sum_{d,m,p,n}D_{m,p,n;d}\chi^d p^m r^p q^n\label{CGauss}
\eea
where $D_{m,p,n;d}$ counts the number of diagrams $R$ with $(n+m+p)$ boxes and $d$ rows, that is the result of 
adding $m+p$ boxes that are randomly distributed over the $d$ rows of the Young diagram $r$ with $n$ boxes. 
Our construction of the Gauss graph operators only holds when the displaced corners approximation holds.
Thus, we trust $D_{m,p,n;d}$ to count the number of Gauss graph operators for a system of $d\sim N^0$ giant gravitons
when $n,m,p\sim N$ and $n\gg m+p$.
This is the main result of this section.

We want to compare this to the counting of ${1\over 8}$-BPS giant gravitons.
As we discussed in the introduction, this counting problem can be mapped to counting energy eigenstates of a system 
of $N$ bosons in a 3-dimensional harmonic oscillator.
The grand canonical partition function for bosons in a 3-dimensional simple harmonic oscillator is given by
\bea
Z(\zeta,q_1,q_2,q_3)=\prod_{n_1=0}^\infty\prod_{n_2=0}^\infty\prod_{n_3=0}^\infty
{1\over 1-\zeta q_1^{n_1}q_2^{n_2}q_3^{n_3}}\label{CBose}
\eea
with the fugacity $\zeta$ being dual to particle number\cite{Mandal:2006tk}.
Notice that (\ref{CGauss}) exactly matches the grand canonical partition function (\ref{CBose}) for bosons in a 
harmonic oscillator potential with $\chi$ playing the role of the fugacity.
This is in harmony with the fact that the number of rows matches the number of giant gravitons.
In the 3-dimensional harmonic oscillator we have 3 types of excitations, counted by $q_1$, $q_2$ and $q_3$.
These map into the three types of boxes ($X$, $Y$ or $Z$ boxes) appearing in $R$, counted by $p$, $q$ and $r$.
Thus, long rows in $R$ map to highly excited particles.
This proves our first claim: the counting of the Gauss graph operators matches the counting of
${1\over 8}$ BPS giant gravitons.

It is straightforward to consider the restriction to the ${1\over 4}$-BPS giant gravitons.
These operators are constructed using only $Z$ and $Y$ fields.
Arguing as above and counting partially labeled Young diagrams with boxes labeled $z$ or $y$, in the displaced
corners approximation, we obtain the generating function
\bea
Z =\prod_{l=0}^\infty\prod_{n=0}^\infty
{1\over 1 -\chi p^l q^n}=
\sum_{d,m,n}D_{m,p,n;d}\chi^d p^m q^n\label{quartergauss}
\eea
This counting can be compared to the counting of ${1\over 4}$-BPS giant gravitons.
This counting problem can be mapped to counting energy eigenstates of a system of $N$ bosons in a 
2-dimensional harmonic oscillator.
The counting (\ref{quartergauss}) does indeed match the grand canonical partition function for bosons in a 
2-dimensional simple harmonic oscillator, which is given by
\bea
Z(\zeta,q_1,q_2,q_3)=\prod_{n_1=0}^\infty\prod_{n_2=0}^\infty
{1\over 1-\zeta q_1^{n_1}q_2^{n_2}}
\eea
Thus, restricting the counting we demonstrates that the counting of the ${1\over 4}$-BPS Gauss graph operators 
matches the counting of ${1\over 4}$ BPS giant gravitons, as it should.

\section{Matching States to Operators}\label{states}

The fact that the number of Gauss graph operators matches the number of energy eigenstates states of a system of
bosons in a 3-dimensional harmonic oscillator potential, motivates us to look for a correspondence between the two.
To start we will consider operators $O^{\vec m,\vec p}_{R,r}(\sigma)$ labeled by Young diagrams that have a single row.
In this case we don't need to encode a complicated shape for $R$, so we will simply list the number of $Z$s,
$Y$s and $X$s in the operator as $O_{n,m,p}$. 
Since this row has $O(N)$ boxes, we have a system of $N$ bosons and one of them is highly excited.
The idea is that since we have one highly excited particle, we can use a single particle description and overlaps of the 
single particle wave functions will match correlation functions of Gauss graph operators in the CFT. 
We focus on $R$'s with a single row because the computations are so simple to carry out in this case that we can
compute many quantities exactly.
There is a simple formula for the Gauss graph operators we consider, in terms of the Schur polynomials
\bea
O_{n,m,p}(Z,Y,X)=
{\cal N} {\rm Tr}\left(Y{d\over dZ}\right)^m
{\rm Tr}\left(X{d\over dZ}\right)^p\chi_{(n+m+p)}(Z)
\eea
where
\bea
{\cal N}=\sqrt{n! (N-1)!\over m!p!(n+m+p)!(N+n+m+p-1)!}
\eea
We are using the notation $(k)$ to denote a Young diagram that has a single row of $k$ boxes.
There are a number of natural operators that act on the Gauss graphs.
For example, we have
\bea
{\rm Tr}(Y{d\over dZ})^{k_1}{\rm Tr}(X{d\over dZ})^{k_2} O_{n,m,p}(Z,Y,X)\,\,\,
\propto\,\,\,
O_{n-k_1-k_2,m+k_1,p+k_2}(Z,Y,X)
\eea
Thus, a natural correlator to consider is given by
\bea
\langle O_{n-k_1-k_2,m+k_1,p+k_2}^\dagger {\rm Tr}\left(Y{d\over dZ}\right)^{k_1}
{\rm Tr}\left(X{d\over dZ}\right)^{k_2}O_{n,m,p}\rangle =
\sqrt{(m+k_1)!(p+k_2)!n!\over m!p!(n-k_1-k_2)!}\label{GGresult}
\eea
To describe a single particle in a 3d harmonic oscillator, we need three sets of creation and annihilation operators
\bea
\big[ a_z,a_z^\dagger \big]=\big[ a_y, a_y^\dagger \big]=\big[ a_x,a_x^\dagger \big]=1
\eea
Using the above oscillators we can create a state with an arbitrary number of $z$ quanta, $y$ quanta or $x$ quanta.
We suggest that the correspondence between Gauss graph operators and particle states is as follows
\bea
O_{n,m,p}\qquad\leftrightarrow\qquad |O_{n,m,p}\rangle ={1\over\sqrt{n! m! p!}}
(a_x^\dagger)^p (a_y^\dagger)^m (a_z^\dagger)^n |0\rangle
\eea
The correspondence identifies the number of $z$, $y$ or $x$ quanta in the particle state with the number of
$Z$s, $Y$s or $X$s in the Gauss graph operator.
There is a natural extension to include operators, suggested by this identification.
For example
\bea
{\rm Tr}(Y{d\over dZ})^{k_1}{\rm Tr}(X{d\over dZ})^{k_2}\qquad \leftrightarrow\qquad 
(a_y^\dagger)^{k_1} (a_x^\dagger)^{k_2} (a_z)^{k_1+k_2}
\eea
As a test of the proposed correspondence, note that
\bea
&&\langle O_{n-k_1-k_2,m+k_1,p+k_2} |(a_y^\dagger)^{k_1} (a_x^\dagger)^{k_2} 
(a_z)^{k_1+k_2} |O_{n,m,p}\rangle \cr\cr
&=&
{\langle 0|(a_z)^{n}(a_y)^{m+k_1}(a_x)^{p+k_2}(a_z^\dagger)^{n}(a_y^\dagger)^{m+k_1}(a_x^\dagger)^{p+k_2}
|0\rangle\over\sqrt{n!m!p!(n-k_1-k_2)!(m+k_1)!(p+k_2)!}}\cr\cr\cr
&=&\sqrt{n!(m+k_1)!(p+k_2)!\over (n-k_1-k_2)! m! p!}
\eea
which is in complete agreement with (\ref{GGresult}). 
Very similar computations comparing, for example
\bea
\langle O^\dagger_{n-k_1,m-k_2,p-k_3}
\left({\rm Tr}{d\over dZ}\right)^{k_1}
\left({\rm Tr}{d\over dY}\right)^{k_2}
\left({\rm Tr}{d\over dX}\right)^{k_3} O_{n,m,p}\rangle
\eea
and
\bea
\langle O_{n-k_1,m-k_2,p-k_3} |(a_z)^{k_1}(a_y)^{k_2} (a_x)^{k_3}|O_{n,m,p}\rangle
\eea
show that we should identify
\bea
a_x\,\,\,\leftrightarrow\,\,\,\sqrt{m+n+p\over N+m+n+p}\,{\rm Tr}\left({d\over dX}\right)\cr
a_y\,\,\,\leftrightarrow\,\,\,\sqrt{m+n+p\over N+m+n+p}\,{\rm Tr}\left({d\over dY}\right)\cr
a_z\,\,\,\leftrightarrow\,\,\,\sqrt{m+n+p\over N+m+n+p}\,{\rm Tr}\left({d\over dZ}\right)
\eea
These computations make use of the reduction rule of \cite{deMelloKoch:2004crq,deMelloKoch:2007rqf}.

We now want to argue that the identifications we have developed above have a natural extension which identifies
Gauss graph operators with $q$ rows with a $q$ particle system.
Towards this end, we first point out a dramatic simplification in the formula for the Gauss graph operators, arising when we
specialize to BPS operators.
As discussed in the introduction, in this case we set the permutation $\sigma$ appearing in (\ref{ggo}) to the identity.
Using the orthogonality of the branching coefficients we then find
\bea
\sum_{j,k}\Gamma^{(s,t)}_{jk}({\bf 1} )B^{(s,t)\to 1_{H_X\times H_Y}}_{j \vec\mu_1}
B^{(s,t)\to 1_{H_X\times H_Y}}_{k \vec\mu_2}
&=&\sum_{j,k}\delta_{jk} B^{(s,t)\to 1_{H_X\times H_Y}}_{j \vec\mu_1}
B^{(s,t)\to 1_{H_X\times H_Y}}_{k \vec\mu_2}\cr
&=&\delta_{\vec\mu_1\vec\mu_2}
\eea
This leads to the following formula (the operators below are normalized to have a unit two point function; they differ
from the operators in (\ref{ggo}) that are not normalized, by a factor of $\sqrt{|H_X\times H_Y|}$)
\bea
O^{\vec m,\vec p}_{R,r}(X,Y,Z)={1\over n! m! p!}\sqrt{|H_X\times H_Y|{\rm hooks}_r\over{\rm hooks}_R f_R}
\sum_{\sigma\in S_{n+m+p}}{\rm Tr}\left(P_{R,r}\Gamma_R(\sigma)\right)
{\rm Tr}(\sigma X^{\otimes p}Y^{\otimes m}Z^{\otimes n})\cr
\eea
$P_{R,r}$ is a projector on the carrier space of $R$.
It projects to the subspace of Young-Yammonouchi states that have $1,2,...,m+p$ distributed in the boxes that
belong to $R$ but not $r$ and $m+p+1,...,m+p+n$ distributed in the boxes that belong to $R$ and $r$.
Using this formula, it is straight forward to prove that
\bea
  {\rm Tr}\left({d\over dX}\right)O^{\vec m,\vec p}_{R,r}(X,Y,Z)
=\sum_{i=1}^q \sqrt{ p_i\,\, c_{RR^{(1)}_i}\over n_i+m_i+p_i}O^{\vec m,\vec p^{(1)}_i}_{R^{(1)}_i,r}(X,Y,Z)\label{Xred}
\eea
\bea
  {\rm Tr}\left({d\over dY}\right)O^{\vec m,\vec p}_{R,r}(X,Y,Z)=\sum_{i=1}^q 
\sqrt{ m_i\,\, c_{RR^{(1)}_i}\over n_i+m_i+p_i}
O^{\vec m^{(1)}_i,\vec p}_{R^{(1)}_i,r}(X,Y,Z)\label{Yred}
\eea
\bea
  {\rm Tr}\left({d\over dZ}\right)O^{\vec m,\vec p}_{R,r}(X,Y,Z)=\sum_{i=1}^q 
     \sqrt{ n_i\,\, c_{RR^{(1)}_i}\over n_i+m_i+p_i}
O^{\vec m,\vec p}_{R^{(1)}_i,r^{(1)}_i}(X,Y,Z)\label{Zred}
\eea
The first formula above is exact.
The last two hold only in the large $N$ limit.
We have introduced some new notation:
the Young diagram $R^{(n)}_i$ is obtained from $R$ by dropping $n$ boxes from row $i$ of $R$.
Further, $\vec p^{(n)}_i$ is obtained from vector $\vec p$ by replacing $p_i\to p_i-n$ and similarly for 
$\vec m^{(n)}_i$.
Finally, $c_{RR^{(1)}_i}$ is the factor of the box that belongs to $R$ but not to $R^{(1)}_i$.
Recall that a box in row $i$ and column $j$ has factor $N-i+j$.
For the proof of these formulas, we use the notation
$$
{\cal N}={1\over n!m!p!}\sqrt{|H_X\times H_Y|{\rm hooks}_r\over{\rm hooks}_R f_R}
$$
and
$$ 
{\rm Tr}(\sigma \, \cdot\, X^{\otimes p}\otimes Y^{\otimes\, m}\otimes Z^{\otimes\, n})
=X^{i_1}_{i_{\sigma (1)}}\cdots X^{i_p}_{i_{\sigma (p)}}Y^{i_{p+1}}_{i_{\sigma (p+1)}}
\cdots Y^{i_{p+m}}_{i_{\sigma (p+m)}}Z^{i_{p+m+1}}_{i_{\sigma (p+m+1)}}\cdots 
Z^{i_{p+m+n}}_{i_{\sigma (p+m+n)}}
$$
We will now prove (\ref{Xred}).
A simple computation shows
\bea
  {dO^{\vec m,\vec p}_{R,r}\over d X^i_i}
&=&p{\cal N}\sum_{\sigma\in S_{n+m+p}}{\rm Tr}(P_{R,r}\Gamma^{(R)}(\sigma)\,)
     {\rm Tr}(\sigma\, \cdot\,1\, X^{\otimes p-1}\otimes Y^{\otimes\, m}\otimes Z^{\otimes\, n})\cr
&=&p{\cal N}\sum_{\sigma\in S_{n+m+p-1}}\sum_{i=1}^{n+m}{\rm Tr}(P_{R,r}\Gamma^{(R)}(\sigma\, (i,1))
{\rm Tr}(\sigma (i,1)\, \cdot\, 1\, X^{\otimes p-1}\otimes Y^{\otimes\, m}\otimes Z^{\otimes\, n})\cr
&=&p{\cal N}\sum_{\sigma\in S_{n+m+p-1}}{\rm Tr}(P_{R,r}\Gamma^{(R)}(\sigma)[N+\sum_{i=2}^{n+m}(i,1)]\,)
{\rm Tr}(\sigma \, \cdot\, 1\, X^{\otimes p-1}\otimes Y^{\otimes\, m}\otimes Z^{\otimes\, n})\nonumber
\eea
Since we are summing over elements of the subgroup $S_{n+m+p-1}\subset S_{n+m+p}$ we can decompose the trace 
over the irreducible representation of $S_{n+m+p}$ as a sum of traces over irreducible representation $R^{(1)}_i$
of the subgroup $S_{n+m+p-1}$.
Now use the fact that $N+\sum_{i=2}^{n+m}(i,1)$ gives $c_{RR^{(1)}_i}=$ the factor of the box dropped from $R$
when acting on any state in the carrier space of $R$ that also belongs to the $R^{(1)}_i$ subspace. 
We find
\bea
{dO^{\vec m,\vec p}_{R,r}\over dX^j_j} = \sum_{i=1}^q \,\, f^{(i)}_{\cal N}\,\, c_{RR^{(1)}_i}
O^{\vec m,\vec p^{(1)}_i}_{R^{(1)}_i,r}
\eea
where the factor
\bea
f^{(i)}_{\cal N}=\sqrt{ p_i \over (n_i+m_i+p_i)c_{RR^{(1)}_i}}
\eea
accounts for the change in the normalization factor ${\cal N}$ of the operator.
This is an exact formula - it does not depend on large $N$ or on the displaced corners approximation.
Next consider the proof of (\ref{Yred}) and (\ref{Zred}).
Consider
\bea
  {dO^{\vec m,\vec p}_{R,r}\over d Y^i_i}
&=&m{\cal N}\sum_{\sigma\in S_{n+m+p}}{\rm Tr}(P_{R,r}\Gamma^{(R)}(\sigma)\,)
     {\rm Tr}(\sigma\, \cdot\, X^{\otimes p}\otimes 1\,\otimes Y^{\otimes\, m-1}\otimes Z^{\otimes\, n})\cr
&=&m{\cal N}\sum_{\sigma\in S_{n+m+p-1}}\sum_{i=1}^{n+m}{\rm Tr}(P_{R,r}\Gamma^{(R)}(\sigma)
{\rm Tr}((p+1,1) \sigma (p+1,1)\, \cdot\, 1\,\otimes\, X^{\otimes p}\otimes Y^{\otimes\, m-1}\otimes Z^{\otimes\, n})\cr
&=&m{\cal N}\sum_{\sigma\in S_{n+m+p-1}}{\rm Tr}(P_{R,r}\Gamma^{(R)}((1,p+1)\sigma)(1,p+1))
{\rm Tr}(\sigma \, \cdot\, 1\,\otimes\, X^{\otimes p}\otimes Y^{\otimes\, m-1}\otimes Z^{\otimes\, n})\nonumber
\eea
The new feature in the above derivation is the presence of the $(1,p+1)\in S_{n+m+p}$ factors needed to swap
the removed $Y$ box to the end of the row so that it can be removed, using the same manipulations as above.
The evaluation of the action of these factors is most easily performed using Young's orthogonal representation, which
gives a rule for the action of adjacent permutations (i.e. permutations of the form $(i,i+1)$) on Young-Yamanouchi 
(hereafter abbreviated YY) states.
Let $|Y\rangle$ denote a YY state, and let $|Y(i\leftrightarrow i+1)\rangle$ denote the YY state obtained by swapping
boxes $i$ and $i+1$.
A box in row $a$ and column $b$ has content given by $b-a$.
Denote the content of the box in $|Y\rangle$ filled with $j$ by $c_j$.
The rule is
\bea
  (i,i+1)|Y\rangle ={1\over c_{i}-c_{i+1}}|Y\rangle +\sqrt{1-{1\over (c_i -c_{i+1})^2}}|Y(i\leftrightarrow i+1)\rangle
\eea
This rule simplifies dramatically in the displaced corners limit, at large $N$.
If the two boxes belong to the same row we find $(i,i+1)|Y\rangle =|Y\rangle$ and  if not
$(i,i+1)|Y\rangle =|Y(i\leftrightarrow i+1)\rangle.$
This is all that is needed to complete the proof of (\ref{Yred}) and (\ref{Zred}) and it proceeds exactly as for the
first rule proved above.
Note that because we used simplifications of the large $N$ limit, (\ref{Yred}) and (\ref{Zred}) are not exact statements
but hold only at large $N$.
The three statements derived above admit some natural generalizations.
For example, we can consider tracing over a product of derivatives to obtain
\bea
  {\rm Tr}\left({d^k\over dX^k}\right)O^{\vec m\vec p}_{R,r}(X,Y,Z)
=\sum_{i=1}^q \left({c_{RR^{(1)}_i}\over n_i+m_i+p_i}\right)^{k\over 2} \,\,\,
\prod_{a=0}^{k-1}\sqrt{p_i-a}\,\,\,
O^{\vec p^{(k)}_i\vec m}_{R^{(k)}_i,r}(X,Y,Z)
\eea
There are obvious generalization when we have a product of $Y$ or $Z$ derivatives.
We could also allow more than one type of derivative in a given trace, for example (in what follows $k=k_1+k_2$)
\bea
&&  {\rm Tr}\left({d^{k_1}\over dX^{k_1}}{d^{k_2}\over dY^{k_2}}\right)O^{\vec m\vec p}_{R,r}(X,Y,Z)\cr
&&=\sum_{i=1}^q \left({c_{RR^{(1)}_i}\over n_i+m_i+p_i}\right)^{k\over 2} \,\,\,
\prod_{a=0}^{k_1-1}\sqrt{p_i-a}\prod_{b=0}^{k_2-1}\sqrt{m_i-b}
\,\,\, O^{\vec m^{(k_2)}_i\vec p^{(k_1)}_i}_{R^{(k)}_i,r}(X,Y,Z)
\eea
By using these formulas for each trace successively, we can also easily evaluate expressions of this form 
\bea
{\rm Tr}\left({d^{k_1}\over dX^{k_1}}{d^{k_2}\over dY^{k_2}}\right)
\cdots  {\rm Tr}\left({d^{k_3}\over dZ^{k_3}}\right)O^{\vec m\vec p}_{R,r}(X,Y,Z)
\eea
To compare to a multi particle system of $q$ noninteracting particles, again in a 3-dimensional harmonic oscillator potential,
we need to introduce $q$ copies of the oscillators ($I,J=1,...,q$)
\bea
\big[ a_z^{(I)},a_z^{(J) \dagger} \big]=\big[ a_y^{(I)}, a_y^{(J) \dagger} \big]
=\big[ a_x^{(I)},a_x^{(J) \dagger} \big]=\delta^{IJ}
\eea
one copy for each particle.
Each Gauss graph operator $O^{\vec m\vec p}_{R,r}$ is specified by giving the number of $Z$ boxes ($r_i$),
$Y$ boxes ($m_i$) and $X$ boxes ($p_i$) in the $i$th row for $i=1,...,q$.
The corresponding multi particle state is
\bea
|O^{\vec m\vec p}_{R,r}\rangle = \prod_{I=1}^q 
{(a_z^{(I)\dagger})^{r_I}\over\sqrt{r_I!}}{(a_y^{(I)\dagger})^{m_I}\over\sqrt{m_I!}}
{(a_x^{(I)\dagger})^{p_I}\over\sqrt{p_I!}}|0\rangle
\eea
Using these formulas we can compare (for example) the matrix elements
\bea
  \langle O^{\vec m\vec p}_{R^{(k)}_q,r^{(k)}_q}|(a_z^{(I)})^k|O^{\vec m\vec p}_{R,r}\rangle
\eea
to the correlation functions
\bea
    \langle O^{\vec m\vec p\,\,\,\dagger}_{R^{(k)}_q,r^{(k)}_q}\, 
   {\rm Tr}\left( {d^k\over dZ^k}\right)\, O^{\vec m\vec p}_{R,r}\rangle
\eea
to learn that we should identify
\bea
{\rm Tr}\left( {d^k\over dZ^k}\right)\leftrightarrow  \sum_{I=1}^q
\left(\sqrt{N+m_I+n_I+p_I\over m_I+n_I+p_I}\right)^k (a_z^{(I)})^k
\eea
In the above formula $n_I$ is the number of $Z$ boxes in row $I$, $m_I$ the number of $Y$ boxes and $p_I$ the
number of $X$ boxes. 
The general rule is ($k=k_1+k_2+k_3$)
\bea
{\rm Tr}\left( {d^{k_1}\over dX^{k_1}}{d^{k_2}\over dY^{k_2}}
{d^{k_3}\over dZ^{k_3}}\right)\leftrightarrow 
\sum_{I=1}^q  \left(\sqrt{N+m_I+n_I+p_I\over m_I+n_I+p_I}\right)^k
(a_x^{(I)})^{k_1} (a_y^{(I)})^{k_2} (a_z^{(I)})^{k_3}
\eea
It is easy to check that the ordering of operators inside the trace on the left hand side above does not matter, when acting 
on the operators we consider, at large $N$.
Multi trace formulas use the above identification for each trace separately.
For example
\bea
&&{\rm Tr}\left( {d^{k_1}\over dX^{k_1}}{d^{k_2}\over dY^{k_2}}
{d^{k_3}\over dZ^{k_3}}\right){\rm Tr}\left( {d^{k_4}\over dX^{k_4}}\right)\cr\cr
&&\quad\leftrightarrow\quad 
\sum_{I=1}^q  \left(\sqrt{N+m_I+n_I+p_I\over m_I+n_I+p_I}\right)^k
(a_x^{(I)})^{k_1} (a_y^{(I)})^{k_2} (a_z^{(I)})^{k_3}\cr\cr
&&\qquad\qquad \times\sum_{J=1}^q  \left(\sqrt{N+m_J+n_J+p_J\over m_J+n_J+p_J}\right)^{k_4}
(a_x^{(I)})^{k_4}
\eea
By comparing overlaps between states with polynomials of creation and annihilation operators sandwiched in between 
and correlators of Gauss graph operators with traces of polynomials of the matrices and derivatives with respect to the 
matrices acting on the Gauss graph operators as in the examples we studied above, we can build any entry in the
dictionary between the $q$ particle system and Gauss graph operators with $q$ rows.

\section{Outlook}\label{outlook}

The description of giant gravitons, constructed using a world volume analysis, allows one to count the
set of all ${1\over 8}$-BPS giant gravitons.
This counting matches $N$ bosons in a 3-dimensional harmonic oscillator.
It is also possible to define an index to count single trace BPS operators, and it can be computed 
both at weak coupling (using the gauge theory) and at strong coupling (as a sum over the spectrum of free massless 
particles in AdS$_5\times$S$^5$).
The results of these different computations are in complete accord.
One can compute the spectrum of anomalous dimensions, for operators with a bare dimension of order $N$,
in the ${\cal N}=4$ super Yang-Mills theory\cite{Carlson:2011hy,Koch:2011hb,deMelloKoch:2012ck}.
In this study we have demonstrated that exactly the same counting (i.e. $N$ bosons in a 3-dimensional harmonic oscillator)
results from counting operators of vanishing anomalous dimension in this spectrum.
Motivated by this agreement, we have looked for a relation between multi particle wave functions and Gauss graph 
operators.
Our basic result is that a map between particle wave functions for particles in a 3-dimensional harmonic oscillator
and Gauss graph operators is easily constructed by comparing overlaps of wave functions of the particle system
with correlators of Gauss graph operators.
The correlator computations have made use of significant simplifications that arise for the BPS Gauss graph operators. 
The number of particles match the number of rows in the Young diagram labeling the Gauss graph operator.
In our opinion, these results provide concrete evidence that the Gauss graph operators are indeed the operators dual
to the ${1\over 8}$-BPS giant gravitons.
To interpret the link between the particle system and the Gauss graph operators, recall the link between giant gravitons 
and an eigenvalue description of the multi matrix dynamics, which has been pursued  
in \cite{Berenstein:2004kk,Berenstein:2013md}. 
Thus, the fact that the matrix model computations appear to be related to the dynamics of non-interacting
particles gives hints as to how matrix model dynamics may simplify, along the line of the proposals 
of \cite{Berenstein:2005aa,Koch:2008ah,Berenstein:2014pma,deMelloKoch:2016whh}. 

Any computation of overlaps performed with our wave functions can be mapped into a computation
of Gauss graph correlators.
However, the wave function picture does clarify the structure of the ${1\over 8}$-BPS operators
in ways that are not manifest in the Gauss graph description.
For example, our wave functions make it clear that a Hilbert space for $N$ 3d bosons emerges from the CFT.
This structure is interesting, as we now explain.
The ${1\over 2}$-BPS sector has a high degree of supersymmetry and so is relatively simple and has often served 
as a bridge connecting the gauge theory and supergravity regimes. 
In the CFT this sector can be consistently decoupled resulting in a system that admits a description in terms of free 
fermions moving in a harmonic oscillator potential. 
This is well understood from the gauge theory point of view where the Lagrangian of the decoupled theory is that
of a complex matrix whose eigenvalues obey Fermi-Dirac statistics, with the statistics induced from the integration measure.
On the gravity side the symplectic form of Type IIB SUGRA encodes the commutation relations that must be imposed 
to quantize the system. 
Restricting this symplectic form to the LLM family of solutions defines a symplectic structure that fixes a quantization 
and ultimately reproduces the free fermion Hilbert space\cite{Grant:2005qc}.
In the same way that free fermion quantum mechanics is equivalent to the singlet sector of a single matrix quantum
mechanics, the wave functions we have written down are equivalent to the BPS Gauss graph operators. 
It would be interesting to recover the Hilbert space of our wave functions by quantizing using the symplectic form of 
Type IIB SUGRA, after restricting to the ${1\over 8}$-BPS family of solutions.

{\vskip 0.5cm}

\noindent
{\it Acknowledgements:}
RdMK would like to thank Ilies Messamah and Sanjaye Ramgoolam for many useful discussions on the topic of this study.
This work is based upon research supported by the South African Research Chairs
Initiative of the Department of Science and Technology and National Research Foundation.
Any opinion, findings and conclusions or recommendations expressed in this material
are those of the authors and therefore the NRF and DST do not accept any liability
with regard thereto.

\end{document}